\documentclass[sigconf]{acmart}

\usepackage{booktabs} 

\usepackage{multirow}
\usepackage{subcaption}
\usepackage{caption}
\usepackage{setspace}
\usepackage{amsmath}
\usepackage[english]{babel}
\definecolor{mygray}{gray}{0.5}
\usepackage{float}
\usepackage{flushend}
\usepackage{mathtools}
\usepackage{soul}
\usepackage{arydshln}            
\usepackage{breqn}
\usepackage{enumitem}

\long\def\comment#1{}
\long\def\comments#1{}

\author{Yifan Qiao}
\affiliation{%
  \institution{Department of Computer Science, University of California}
  \city{Santa Barbara}
  \state{California}
  \postcode{93106}
  \country{USA}
}

\author{Yingrui Yang}
\affiliation{%
  \institution{Department of Computer Science, University of California}
  \city{Santa Barbara}
  \state{California}
  \postcode{93106}
  \country{USA}
}

\author{Shanxiu He}
\affiliation{%
  \institution{Department of Computer Science, University of California}
  \city{Santa Barbara}
  \state{California}
  \postcode{93106}
  \country{USA}
}

\author{Tao Yang}
\affiliation{%
  \institution{Department of Computer Science, University of California}
  \city{Santa Barbara}
  \state{California}
  \postcode{93106}
  \country{USA}
}



\AtBeginDocument{}

\copyrightyear{2023}
\acmYear{2023}
\setcopyright{rightsretained}
\acmConference[SIGIR '23]{Proceedings of the 46th International ACM SIGIR Conference on Research and Development in Information Retrieval}{July 23--27, 2023}{Taipei, Taiwan}
\acmBooktitle{Proceedings of the 46th International ACM SIGIR Conference on Research and Development in Information Retrieval (SIGIR '23), July 23--27, 2023, Taipei, Taiwan}\acmDOI{10.1145/3539618.3592051}
\acmISBN{978-1-4503-9408-6/23/07}

\makeatletter
\gdef\@copyrightpermission{
 \begin{minipage}{0.3\columnwidth}
  \href{https://creativecommons.org/licenses/by/4.0/}{\includegraphics[width=0.90\textwidth]{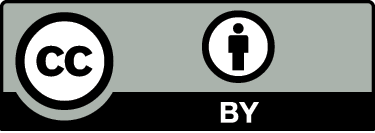}}
 \end{minipage}\hfill
 \begin{minipage}{0.7\columnwidth}
  \href{https://creativecommons.org/licenses/by/4.0/}{This work is licensed under a Creative Commons Attribution International 4.0 License.}
 \end{minipage}
 \vspace{5pt}
}
\makeatother

\begin{document}

\title{ Representation Sparsification with Hybrid Thresholding\\ for Fast SPLADE-based  Document Retrieval}

\begin{abstract}

Learned sparse  document representations using a transformer-based neural model
has been found to be  attractive in both relevance effectiveness and time efficiency.
This paper describes a representation sparsification scheme  based on hard and soft thresholding with an inverted index approximation 
for faster SPLADE-based document retrieval.
It provides analytical and  experimental results on  the impact of this learnable hybrid thresholding scheme.


\end{abstract}

\maketitle

\section{Introduction}

\comments{
The retrieval stage of large-scale search often uses
a sparse data structure called an 
inverted index~\cite{Ricardo2011} that  contains a set of tokens such as keywords or  terms.
Each  of them points to a list of document IDs that possess a feature for such a token.
a {\em posting} record.
A popular method for document retrieval with an inverted index is to use 
BM25 based additive ranking and  term frequency~\cite{Robertson2009BM25}.
}

Recently learned sparse retrieval techniques~\cite{Zamani2018SNRM,
Dai2020deepct, Mallia2021deepimpact, Lin2021unicoil,2021NAACL-Gao-COIL,
Formal2021SPLADE, Formal2021SPLADEV2,Formal_etal_SIGIR2022_splade++}
have become attractive
because  such a representation  can deliver  a strong  relevance 
by leveraging transformer-based models to expand  document tokens with learned weights
and can  take advantage of traditional inverted index based retrieval techniques~\cite{Mallia2017VBMW,mallia2022faster}. 
Its query processing is cheaper than a dense representation which requires GPU support
(e.g. ~\cite{xiong2021-ANCE, 2021EMNLP-Ren-RocketQAv2,  Santhanam2021ColBERTv2})
even with efficiency optimization  through approximate nearest neighbor  
search~\cite{johnson2019billion, 2022WSDM-Zhan-RepCONC, Xiao2022Distill-VQ}.

This paper focuses on the SPLADE family of sparse 
representations~\cite{ Formal2021SPLADE, Formal2021SPLADEV2,Formal_etal_SIGIR2022_splade++}
 because it can deliver
a high MRR@10 score for MS MARCO passage ranking~\cite{Craswell2020OverviewOT} and
a strong zero-shot performance for the BEIR datasets~\cite{thakur2021beir}, which are well-recognized IR benchmarks.
The sparsification optimization in SPLADE has used L1 and FLOPS regularization to minimize non-zero weights  
during model learning, and our objective is to exploit additional opportunities to further increase the sparsity of 
inverted indices produced by SPLADE.
Earlier static inverted index pruning research ~\cite{2001SIGIR-Carmel,2007SIGIR-Blanco, 2006CIKM-Buttcher} for a lexical model has shown 
the usefulness of trimming a term posting list or a document by a limit. 
Yang et al.~\cite{2021-Yang-Masking} conduct  top token  masking by limiting the top activated weight count uniformly
per document
and gradually reducing this weight count limit  to a targeted constant during training.  
Motivated by these studies~\cite{2001SIGIR-Carmel,2007SIGIR-Blanco, 2006CIKM-Buttcher,2021-Yang-Masking} and 
since  they have not addressed  the learnability of a pruning limit through relevance-driven training,
this paper exploits a learnable thresholding architecture
to filter out unimportant neural weights produced by the SPLADE model through joint training. 
\comments{
Learnable thresholding  has been investigated in  the deep learning  community 
for  model pruning, e.g.  
parameter sparsification~\cite{kusupati2020soft,park2020dynamic,LIU2020Dynamic}
and activation sparsity~\cite{2019CVPR-activationmap, 2020ICML-Kurtz-ActiviationSparsity}.
Such techniques have not been studied in the context of sparse retrieval, and 
the impact of thresholding  on relevance and query processing time with inverted indices,  
requires new design considerations and model structuring for document retrieval, even the previous work can be leveraged. 
}

The contribution of this paper is 
a learnable hybrid hard and soft thresholding scheme with an inverted index approximation to
increase  the sparsity of SPLADE-based document and query feature vectors for faster retrieval. 
In addition to experimental validation with MS MARCO and BEIR datasets,
we provide an analysis of the impact of hybrid thresholding with joint training on
index approximation errors and training update effectiveness.

\comments{

\citet{Dai2020deepct} learns  contextualized term weights to replace TF-IDF weights.
}

\section{Background}

For a query $q$ and a document $d$, after expansion and encoding, they can be represented by vector 
$\vec{w}(q)$ and $\vec{w}(d)$ with length $|V|$, 
where $V$ is the vocabulary set. 
The rank score of $q$ and $d$ is computed as
$R(q, d) = \vec{w}(q) \cdot \vec{w}(d) = \sum_{i=1}^{|V|} w_i^q \times w_i^d$.
For sparse vectors with many zeros, retrieval can utilize 
a data structure called inverted index during online inference for fast score 
computation ~\cite{Mallia2017VBMW,mallia2022faster}. 
The SPLADE model uses the BERT token space to predict the feature vector $\vec{w}$. 
In its latest SPLADE++ model, it first calculates the importance of $i$-th input token in $d$ for 
each $j$ in $V$:
$w_{ij}(\Theta)  = {\rm Transform}(\vec{h_i})^T \vec{E_j}+b_j,$
where $\vec{h_i}$ is the BERT embedding of $i$-th token in $d$,
$\vec{E_j}$ is the BERT input embedding for $j$-th token.  
Transform() is a linear layer with GeLU activation and LayerNorm.
The weights in this linear layer, $\vec{E_j}$, 
 and $b_j$ are the SPLADE parameters updated during training and we call them set  $\Theta$.
Then the $j$-th entry $w_j$  of document $d$ (or a query)  is max-pooled as
$w_j(\Theta) =\max_{i\in d} \{ \log(1+{\rm ReLU}(w_{ij} (\Theta))) \}.$
Notice that $w_j \geq 0$.

\comments{
The feature vector of the query $q$ is computed in the same way, 
and the ranking score of a query $q$ and 
a document $d$ is calculated as $\vec{w}(q) \cdot \vec{w}(d)$. 
In order to keep the $\vec{w}$ sparse, the FLOPS loss is jointly optimized with the MarginMSE ranking loss.
}

The loss function of SPLADE models~\cite{ Formal2021SPLADE, Formal2021SPLADEV2,Formal_etal_SIGIR2022_splade++}
contains a per-query ranking loss $L_R$ and sparsity regularization.
The ranking loss has evolved from a log likelihood based function  for maximizing positive document probability
to margin MSE for knowledge distillation. 
This paper uses the loss of SPLADE with a combination that delivers the best result in our training process.
$L_{R}$ is the ranking loss with margin MSE for knowledge distillation~\cite{Hofsttter2020marginMSE}.
The document token regularization $L_D$ is computed on the training documents in each batch based on FLOPS regularization. 
The query token regularization  $L_Q$ is based on L1 norm. 
Let $B$ be a set of training queries with $N$ documents involved in a batch.
$
L_Q = 
\sum_{j \in V} \frac{1}{|B|}\sum_{q \in B}  w_j^q;$  
$ L_D =\sum_{j \in V}( \frac{1}{N} \sum_{ d =1}^{N} w_j^d  )^2.
$

{\bf Related work}.
\comments{
{\bf Learned sparse representation.} 
Recent work on learned sparse representation includes
SPLADE~\cite{Formal2021SPLADE, Formal2021SPLADEV2} learning token importance for  document expansion with sparsity control.
DeepImpact~\cite{Mallia2021deepimpact} learns neural term weights on documents expanded by DocT5Query~\cite{Cheriton2019doct5query}.
COIL~\citep{2021NAACL-Gao-COIL} proposed a contextualized exact match retrieval architecture by storing contextualized token weight vectors in inverted lists.
Since COIL's vector space need can be expensive,  uniCOIL~\citep{Lin2021unicoil} coverts the  vector weights of COIL
into scalar term weights after neural document expansion.
Document retrieval using learned sparse  representations have shown to deliver strong relevance results while retaining a good efficiency based on inverted indices.

\subsection{Efficient Sparse Representation}
}
Other than SPLADE, sparse retrieval studies include
SNRM ~\cite{Zamani2018SNRM},
DeepCT~\cite{Dai2020deepct},
DeepImpact~\cite{Mallia2021deepimpact}, and uniCOIL~\cite{Lin2021unicoil,2021NAACL-Gao-COIL}.
The sparsity of a neural network is studied in  the deep learning  community. 
Soft thresholding in \cite{kusupati2020soft} 
adopts a learnable threshold with function $S(x,t)=ReLU(x-t)$ to make parameter $x$ zero under threshold $t$.
A hard thresholding function $H(x,t)=x$ when $x \ge t$ otherwise 0.
Approximate hard thresholding~\cite{park2020dynamic} uses a Gauss error function to approximate $H(x,t)$ with smooth
gradients.
Dynamic sparse training~\cite{LIU2020Dynamic} finds a dynamic threshold with marked layers.
These works including the recent ones~\cite{2022ICML-pruning-SPDY}
are targeted for sparsification of parameter  edges in  a deep neural network.
In our context, a token weight $w_j$  is an output node in a network. 
The sparsification of output nodes 
is addressed in activation map compression~\cite{2019CVPR-activationmap} 
using ReLU as soft thresholding together with L1 regularization.
The work of ~\cite{2020ICML-Kurtz-ActiviationSparsity} further boosts
sparsity with the Hoyer regularization and a variant of ReLU.
The above techniques have not been investigated in the context of sparse retrieval, and
the impact of thresholding  on relevance and query processing time with inverted indices,
requires new design considerations and model structuring for document retrieval, even the previous work can be leveraged.

\section{Hybrid Thresholding (HT)}
\vspace{-0.2cm}
\begin{figure}[h]
    \centering
    \includegraphics[width=0.8\linewidth]{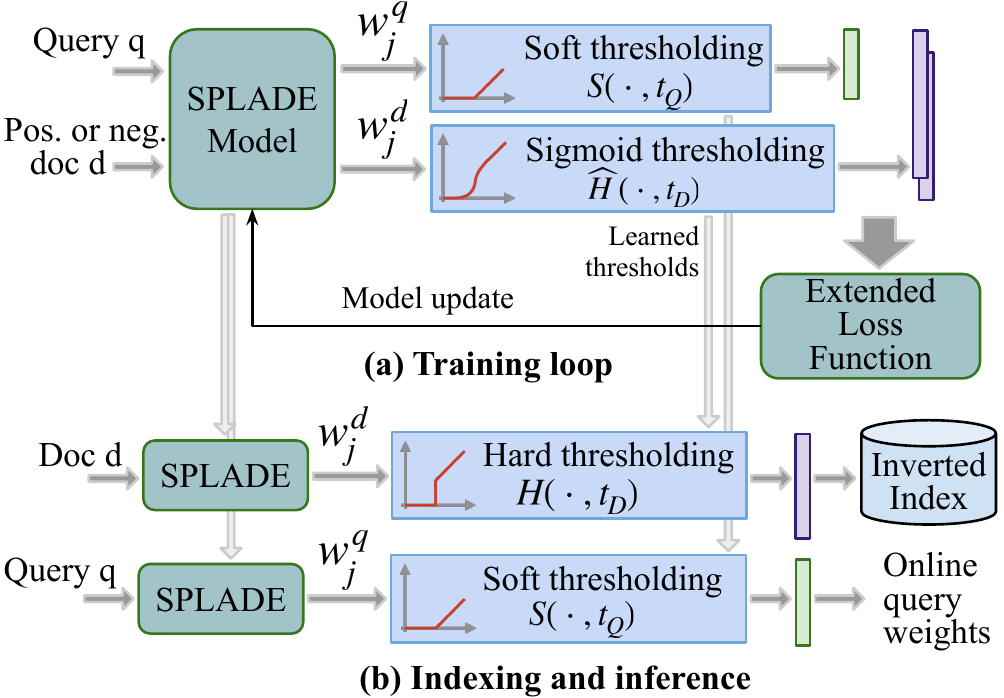}
    \caption{Hybrid  thresholding with an index approximation}
    \label{fig:thresholding} 
\end{figure}

\vspace{-0.2cm}
\comments{
\subsection{Problem Definition}

For a query $q$ and a document $d$, after expansion and encoding, they can be represented by vector $\mathbf{w}(q)$ and $\mathbf{w}(d)$ with length $|V|$, where $V$ is the vocabulary set. The relevance score of $q$ and $d$ is computed as
$$rel(q, d) = \mathbf{w}(q) \cdot \mathbf{w}(d) = \sum_{i=0}^{|V|-1} w_i(q) \times w_i(d)$$
and $w_i \geq 0$.
The sparsity is measured by $\lVert\mathbf{w}(q)\rVert_0$ and $\lVert\mathbf{w}(d)\rVert_0$. We want to 
minimize them while preserving the ranking performance. 
}

{\bf Design considerations}.
To zero out  
a token weight below a learnable threshold, 
there are two options: soft thresholding~\cite{kusupati2020soft}, 
and approximate hard thresholding~\cite{park2020dynamic}. 
\comments{
As discussed below, there is a challenge that  the subgradient for  hard thresholding is always 0  
and thus an approximation needs to be carried out.
We will conduct a joint training of the SPLADE model for both online inference and neural weight derivation 
to find the best thresholds for document tokens and query tokens.
There are two options to map token weights using threshold: soft thresholding and hard thresholding.

Since document token weights  use an approximate hard thresholding function,
the actual weights used during training are not exactly zero in many cases. Our interference uses the exact hard thresholding function, 
and thus there is a gap between weights trained and 0 weight used. Our design goal is to minimize the gap between them. 
}
For query token weights, we find that soft thresholding does not affect relevance significantly.
For document token weights, our study finds that 
compared to soft thresholding, 
hard thresholding 
can retain relevance better 
since it does not change token weights when exceeding a threshold. 
Since the subgradient for  hard thresholding with respect to a threshold is always 0,
 an approximation needs to be carried out
for training. 
For search index generation, an inverted index  produced with the same approximate hard thresholding  as training
keeps many  unnecessary non-zero document token weights, slowing down retrieval significantly.
Thus  we directly apply hard thresholding with a threshold learned from training,
as shown in Figure~\ref{fig:thresholding}.
There is a gap between trained document token weights and actual weights used in our inverted index generation
and online inference, and we intend to minimize this gap (called an index approximation error). 

Thus our design takes a hybrid approach that applies soft thresholding to query token weights during training and inference
and applies approximate hard thresholding to  document token weights during training while using hard thresholding for documents
during index
generation.  
For approximate hard thresholding,
we propose to use a logistic sigmoid-based function instead of a Gauss error function~\cite{park2020dynamic}. 
This sigmoid thresholding simplifies our analysis
of the impact of its hyperparameter choice to 
index approximation errors, and to training stability.
\subsection{Trainable and approximate thresholding }

\comments{
Our approach is designed as follows.
 applying a thresholding function 
 to query tokens is to use a soft thresholding function since it sub-gradients are computable.
while thresholding for document tokens is based post processing with hard thresholding approximation.
}

Training computes  threshold parameters $t_D$, and $t_Q$ for documents and queries, respectively.
From the output of the SPLADE model,
every token weight of a query is replaced with $S(w_j^q, t_Q)$, which is $ReLU(w_j^q-t_Q)$,
and  every document token weight is replaced with $\widehat{H} (w_j^q, t_D)$ 
before their dot product is computed during training as shown in Figure~\ref{fig:thresholding}(a).
Sigmoid thresholding  $\widehat{H}$ is defined as: 
\begin{equation}
{\rm \widehat{H}}(w_j^d,t_D) = w_j^d \sigma(K(w_j^d - t_D)) \mbox { where }
\sigma(x) = \frac{1}{1+ e^{-x}}.
\end{equation}
Here $K$ is a hyperparameter to control the slope steepness of step approximation that jumps from 0 to 1 when exceeding a threshold.

The indexing process uses hard thresholding to replace all document weights  that are below threshold $t_D$ as 0
as depicted  in Figure~\ref{fig:thresholding}(b). 
The above post processing introduces an index approximation
 error $E=|\widehat{H}(w_j^d,t_D) - H(w_j^d, t_D)|$.
We derive its upper bound as follows. Notice that $w_j \ge0$, and
for any $x\ge 0$, $1+x \leq e^{x}$. 
\[
E= w_j^d \sigma(K(w_j^d-t_D))
=\frac{w_j^d}{ 1+ e^{K(t_D-w_j^d)}}
\leq  \frac{w_j^d}{2+ K(t_D-w_j^d)}.
\]
When $w_j^d \ge  t_D$, we can derive  that
\[
E=  w_j^d (1- \sigma(K(w_j^d-t_D)))
=  w_j^d  \sigma(K(t_D-w_j^d)))
\leq  \frac{w_j^d}{2+ K(w_j^d-t_D)}.
\]
Let $\sigma^-$ denote $\sigma(K(w_i^d-t_D))$. $0 < \sigma^- <1$. 
In both of the above cases, the error upper bound is minimized  when $K$ is large.
This is consistent with the fact that error $E$ is monotonically decreasing as $K$ increases 
because $\frac{\partial E}{\partial K}= - w_j^d$ 
$\sigma^- (1-\sigma^-)$ 
$|w_j^d -t_D| \leq 0$. 
When $|w_j^d-t_D|$ is big, the error is negligible
and when $|w_j^d-t_D|$ is small, the error  could become big with a small $K$ value.
But as shown later, an excessively large $K$ value could cause a big parameter update during a training step, affecting joint
training stability. 

Let $Dlen$ and $Qlen$ be the  non-zero token weight count  of  document $d$ and query $q$, respectively.
For our hybrid thresholding,
$
Dlen=\sum_{j}  \textbf{1}_{w_j^d \ge t_D}, 
Qlen=\sum_{j}  \textbf{1}_{w_j^q \ge t_Q}.
$
Here $\textbf{1}_{x \ge y}$ is an indicator function as 1 if $x \ge y$ otherwise 0. 
When increasing $t_D$ and $t_Q$, $Dlen$ and $Qlen$  decrease. 
Thus for a batch of training queries $B$, the original SPLADE loss is extended as: $L= (\frac{1}{|B|}\sum_{q \in B} L_R)
+\lambda_Q L_Q + \lambda_DL_D + \lambda_T L_T.$
The extra item added  is $L_T = \log(1 + e^{-t_D}) + \log(1 + e^{-t_Q})$.
We retain the original $L_Q$ and $L_D$ expressions because 
as $w_j^q$ or $w_j^d$ decreases, more weights can quickly be zeroed out.


\subsection{Threshold and token weight updating}

\comments{
The previous  sparsification studies in neural models in classification (e.g. ~\cite{park2020dynamic,LIU2020Dynamic})
typically use of  one of the thresholding loss regularization 
L1 or FLOPS based weight regularization. 
We keep  both because the term and query  weights in the SPLADE model 
are output nodes instead of parameters of the neural network trained. 
We justify the benefit of this approach by   
}
We study  the change of  $t_D$, $t_Q$, $w_j^d$, and $w_j^q$  
after each training step with
a mini-batch gradient descent update.
The analysis below uses  the first-order Taylor polynomial approximation
and follows the fact that sigmoid  thresholding $\widehat{H}$ and soft thresholding function $S$
are used  independently for a query and a document in the loss function. 
Symbol  $\alpha$ is the learning rate.  
Let ``$d \triangleleft q$'' mean $d$ is  a positive or negative document of query $q$.

\begin{dmath*}
    \Delta t_D =
t^{new}_D -t^{old}_D 
\hiderel{=} -\alpha \frac{\partial L}{\partial t_D} \hiderel{=}
-\alpha \left(
 \frac{1}{|B|}\sum_{q \in B} \frac{\partial L_R}{\partial \widehat{H}} 
 \frac{\partial \widehat{H} }{\partial t_D}  +
\lambda_{T}  \frac{\partial L_T}{\partial t_D}\right) \\
 = \alpha \left(
 \frac{1}{|B|}\sum_{q\in B} \left( 
K
\frac{\partial L_R}{
\partial \widehat{H}} 
\sum_{ d \triangleleft q} \sum_i w_i^d (1-\sigma)^-  \sigma^-
\right)  + 
\lambda_{T} \frac{e^{-t_D}}{1+e^{-t_D}}\right).
\end{dmath*}

\vspace{-0.2cm}

\begin{dmath*}
\Delta t_Q =
t_Q^{new} -t_Q^{old}  
\hiderel{=} -\alpha \frac{\partial L}{\partial t_Q} \hiderel{=}
-\alpha \left(
\frac{1}{|B|}\sum_{q \in B}  \frac{\partial L_R}{\partial S} 
 \frac{\partial S }{\partial t_Q}  +
\lambda_{T}  \frac{\partial L_T}{\partial t_Q}\right)
= \alpha \left(
 \frac{1}{|B|}\sum_{q\in B} \left( 
\frac{\partial L_R}{\partial S} 
 \sum_{i }   \textbf{1}_{w_i^q \ge t_Q} 
\right) + 
\lambda_{T} \frac{e^{-t_Q}}{1+e^{-t_Q}}\right).
\end{dmath*}

\vspace{-0.2cm}
\begin{dmath*}
\Delta w_j^d =
w_{j}^{d,new} -w_{j}^{d,old}  
\hiderel{\approx}  \sum_{\theta \in \Theta}  \frac{\partial w_j^d}{\partial \theta} \Delta \theta
\hiderel{=}-\sum_{\theta \in \Theta}  \frac{\partial w_j^d}{\partial \theta} 
  \alpha  \frac{\partial L}{\partial \theta} 
=-\alpha 
\sum_{\theta \in \Theta}  \frac{\partial w_j^d}{\partial \theta} 
  \left(
 \frac{1}{|B|}\sum_{q\in B} \left(
   \frac{\partial L_R}{\partial \widehat{H}}  
   \left(
   \sum_{d \triangleleft q} \sum_{i}
   \frac{\partial \widehat{H}}{\partial w_i^d}  
   \frac{\partial w_i^d}{\partial \theta} 
    \right)
	+ \\
 \frac{\partial L_R}{\partial S} 
   \left(
   \sum_{i}
\frac{\partial S}{\partial w_i^q} 
   \frac{\partial w_i^q}{\partial \theta} 
    \right)
\right)
+ \lambda_D 
\frac{\partial L_D} {\partial \theta} 
+ \lambda_Q 
\frac{\partial L_Q} {\partial \theta} 
\right).
\end{dmath*}
Notice that  $\frac{\partial \widehat{H} }{\partial w_i^d}  
    = \sigma^-  + K w_i^d \sigma^- (1-\sigma^- )$. 
\comments{
\[
=-\alpha 
\sum_{\theta \in \Theta}   \frac{\partial w_j}{\partial \theta} 
\sum_i \frac{\partial w_i}{\partial \theta} 
  \left( \frac{\partial L_R}{\partial \widehat{H}}  
    (\sigma  + K w_i \sigma (K(w_i-t_D)(1-\sigma (K(w_i-t_D))))
+\frac{\partial L_R}{\partial S}  
 1_{w_i>t} 
 + \lambda_Q \frac{\partial L_Q}{\partial w_i}  
+ \lambda_D \frac{\partial L_D}{\partial w_j}  
\right)
\]}
The above results indicate:
\begin{itemize}[leftmargin=*]
\item A significant number of terms in $\Delta t_D$ and 
$\Delta w_j^d$ 
involve linear coefficient $K$. 
This is verifiably  true also for 
$\Delta w_j^q$. 
Although a large $K$ value can minimize the index approximation error
 $|\widehat{H}(w_j^d,t_D) - H(w_j^d, t_D)|$,  
it can 
cause an aggressive change of token weights and thresholds 
at a training iteration, making training overshoot and miss the global optimum.
Thus  $K$ cannot be too large, and 
our evaluation  further studies this.
\item If $\frac{\partial L_R}{\partial \widehat{H}} \ge 0$,
$\Delta t_D \geq 0$, and the document threshold increases,
decreasing $Dlen$.
Otherwise document token threshold may decrease after a parameter update step during training,  
and the degree of decreasing is reduced by a positive value $\frac{e^{-t_D}}{1+e^{-t_D}}$.
Based on the sign of  $\frac{\partial L_R}{\partial S}$, we can draw a similar conclusion
on 
$\Delta t_Q$.
\end{itemize}


\section{Evaluation}

Our evaluation uses MS MARCO passages~\cite{Craswell2020OverviewOT}
and BEIR datasets~\cite{thakur2021beir}. 
MS MARCO has  8.8M passages while BEIR has 13 different datasets of varying sizes up-to 5.4M.
As a common practice, 
we report the relevance in terms of 
mean reciprocal rank MRR@10 for the 
MS MARCO passage Dev query set with 6980 queries, and the normalized discounted cumulative 
gain nDCG@10~\cite{NDCG} for its DL'19 and DL'20 sets, and also for BEIR.
For retrieval with a SPLADE inverted index,
we report the mean response time (MRT) and 99th percentile time ($P_{99}$) in milliseconds.
The query encoding time is not included.
For the SPLADE model, we warm up it following~\cite{Formal_etal_SIGIR2022_splade++, lassance2022efficiency},
and train it with $\lambda_Q=0.01$ and $\lambda_D=0.008$, and hybrid thresholding.
We use the PISA~\cite{mallia2019pisa} search system to index documents and search queries 
using SIMD-BP128 compression~\cite{2015Lemire} and MaxScore retrieval~\cite{2019ECIRMallia, mallia2022faster}.
Our evaluation runs as a single thread on a Linux CPU-only  server with Intel i5-8259U 2.3GHz  and 32GB memory.
Similar retrieval latency results are observed on a 2.3GHz AMD EPYC 7742 processor.
The checkpoints and related code will be released in https://github.com/Qiaoyf96/HT.

\vspace{-0.1cm}

\begin{table}[tpbh]
    \centering
        \caption{Overall results on MS MARCO passages}
\label{tab:overall}
    \vspace{-0.2cm}
\resizebox{1.0\columnwidth}{!}{%
    \begin{tabular}{l|ccc|cc|r}
    \hline
     \textbf{Methods} & \textbf{MRR} & \textbf{MRT($P_{99}$)} & \textbf{MRT($P_{99}$)} & \textbf{nDCG} & \textbf{nDCG} & \textbf{Dlen} \\ 
      & Dev & \textbf{$k=10$} & \textbf{$k=1000$} & DL'19 & DL'20 \\ 
     \hline 
    SPLADE & \textbf{0.3966}  & 48.3(228) & 127(408) & \textbf{0.7398} & \textbf{0.7340} & 351 \\
    \ \ \ /DT~\cite{park2020dynamic} & 0.3922 & 102(457) & 262(786) & 0.7392 & 0.7319 & 444 \\
    \ \ \ /Top305~\cite{2021-Yang-Masking} & 0.3962 & 42.4(202) & 114(369) & 0.7353 & 0.7288 & 277 \\
    \ \ \ /Top100~\cite{2021-Yang-Masking} & 0.3908 & 21.8(106) & 62.5(196) & 0.7192 & 0.7119 & 99 \\
    \ \ \ /DCP50\%~\cite{2006CIKM-Buttcher} & 0.3958 & 30.0(145) & 83.9(271) & 0.7385 & 0.7321 & 175 \\
    \ \ \ /DCP40\%~\cite{2006CIKM-Buttcher}  & 0.3933 & 25.9(124) & 73.3(235) & 0.7335 & 0.7280 & 140 \\
    \ \ \ /DCP30\%~\cite{2006CIKM-Buttcher}  & 0.3912 & 21.6(101)  & 61.8(193) & 0.7287 & 0.7217 & 105 \\
    \hline
    \ \ \ /Cut0.5 & 0.3924  & 21.9(104) & 62.6(195) & 0.7296 & 0.7212 & 144 \\
    \ \ \ /Cut0.8 & 0.3885 & 15.6(70.4) & 43.8(128) & 0.7207 & 0.7118 & 112 \\

    \hline
    \ \ \ /HT$_1$ & 0.3955 & 22.8(108) & 62.3(195) & 0.7322 & 0.7210 & 140\\

    \ \ \ /HT$_3$ & 0.3942 & 14.2(67.2) & 40.6(123) & 0.7327 & 0.7228 & 106 \\
     \hline

    \ \ \ /HT$_1$-2GTI~\cite{20232GT} & 0.3959 & 10.0(49.1) & 27.6(92.2) & 0.7330 & 0.7210 & 140\\
    \ \ \ /HT$_3$-2GTI~\cite{20232GT} & 0.3942 & \ \ \textbf{6.9(33.9)} & \textbf{19.3(62.1)} & 0.7320 & 0.7228 & 106\\
    \hline
     
    \end{tabular}
   }
\end{table}

\vspace{-0.2cm}

\textbf{Overall results with MS MARCO.} 
Table \ref{tab:overall}
is   a comparison with the baselines  on MS MARCO passage Dev set, DL'19, and DL'20. 
It lists the average $Dlen$ value, and  top-$k$ retrieval time with depth $k=10$ and 1000. 
Row 3 is for original SPLADE trained by ourselves with 
an MRR number  higher than 0.38 reported in ~\cite{Formal_etal_SIGIR2022_splade++,lassance2022efficiency}.
Rows 12 and 13  list the result of our hybrid thresholding marked as HT$_{\lambda_T}$ 
and $K=25$. 
With $\lambda_T=1$, SPLADE/HT$_1$ converges to a point where $t_Q=0.4$ and $t_D=0.5$,  which is  about 2x faster in retrieval.
HT$_{3}$ with $\lambda_T=3$ converges at $t_Q=0.7$ and $t_D=0.8$, resulting 3.1x speedup than SPLADE while having 
a slightly lower MRR@10 0.3942. No statistically significant degradation in relevance
has been observed at the 95\% confidence level for both HT$_1$ and HT$_3$.
The inverted index size reduces from 6.4GB for original SPLADE to 2.8GB and 2.2GB for HT$_1$ and HT$_3$ respectively.
\comments{
The dynamic thresholding~\cite{park2020dynamic} shown as DT in Table \ref{tab:overall} applies hard 
thresholding during training and inference, and the threshold $t$ ends up 0.8 after 20 epochs, 
while still rising. Without regularization for $w$, the sparsity cannot be well controlled, and the retrieval latency is even longer. 
We also follow the top-$k$ masking~\cite{2021-Yang-Masking} and list two configurations as 
baselines in Table \ref{tab:overall}: (1) if the top 305 tokens as in \cite{2021-Yang-Masking} are used for 
representation of each query or document, the model achieves almost the same relevance and similar retrieval latency 
compared with the SPLADE model without thresholding; (2) if the model is continued to be trained and the number of tokens 
used for each representation shrinks to 100, more than 2x speedup is observed, but the relevance drops to 0.3908 MRR@10.
}
When applying two-level guided traversal 2GTI~\cite{20232GT} with its fast configuration, 
Rows 14 and 15 show a further latency reduction to 6.9ms or 19.3ms.

We discuss other baselines listed in this table.
Row 4 named DT uses the thresholding scheme from ~\cite{park2020dynamic}.
Its training does not converge  
with its loss function, 
and  its retrieval is much slower. 
Rows 5 and 6 follow  joint training of top-$k$ masking~\cite{2021-Yang-Masking} 
with the top 305 tokens as suggested  in \cite{2021-Yang-Masking} 
and with the top 100 tokens. 
Rows 7, 8 and 9 marked with DCP$x$ 
follow document centric pruning~\cite{2006CIKM-Buttcher}
that keeps $x$ of top tokens per document where $x$=50\%, 40\%, and  30\%.
We did not list term centric pruning~\cite{2001SIGIR-Carmel,2007SIGIR-Blanco}
because ~\cite{2006CIKM-Buttcher} shows DCP is slightly better in relevance under the same latency constraint.
Rows 10 and 11 with ``/Cut0.5'' and ``/Cut0.8'' apply a hard threshold with 0.5 and 0.8 in
the output of original SPLADE without joint training.
The index pruning options without learning  
from Rows 5 to 11 
can either reduce the latency to  the same level as HT, but their relevance score is  visibly lower;
or  have a relevance similar  to HT  but with much slower latency.
This illustrates the advantage of learned hybrid thresholding with joint training.
 
\comments{
 propose two configurations of our proposed hybrid thresholding HT$_{\lambda_T}$ with different $\lambda_T$. With $\lambda_T=1$, the model converges at a point where $t_q=0.4$ and $t_d=0.5$. This model has the similar latency speedup with top-100 masking, but the 0.3955 MRR@10 relevance is much higher (0.3955 vs. 0.3908). If a more aggressive $\lambda_T=3$ is used, the model converges at $t_q=0.7$ and $t_d=0.8$, resulting 3.1x speedup, with 0.3942 MRR@10. To further illustrate the effectiveness of joint training, we try to apply these thresholds to post-process the SPLADE model. The ``/cut0.5'' and ``/cut0.8'' sets all the $w_d \leq 0.5$ or $w_d \leq 0.8$ to be 0. The same level of latency compared with HT can be observed, but the relevance drops to 0.3924 and 0.3885 MRR@10 respectively.
}


\begin{table}[htpb]
\small
    \centering.
    \caption{Zero-shot performance on BEIR datasets}
            \label{tab:beir}
        \vspace{-0.2cm}

        \resizebox{0.9\columnwidth}{!}{%
    \begin{tabular}{l|cr|cr|cr}
    \hline
        & \multicolumn{2}{c|}{SPLADE} & \multicolumn{2}{c|}{SPLADE/HT$_1$} & \multicolumn{2}{c}{SPLADE/HT$_3$} \\
\textbf{Dataset} & \textbf{nDCG} & \textbf{MRT} & \textbf{nDCG} & \textbf{MRT} & \textbf{nDCG} & \textbf{MRT}\\
 \hline 
               DBPedia  & 0.430 & 135 & \textbf{0.435} & 64.2 & 0.426 & 32.3 \\
        FiQA & \textbf{0.354} & 6.5 & 0.345 & 4.0 & 0.336  & 3.2  \\
        NQ  & \textbf{0.547} & 81.8  & 0.545 & 45.9 & 0.539 & 28.6 \\
        HotpotQA  & 0.678 & 481 & \textbf{0.680} & 265 & 0.678 & 140 \\
        NFCorpus  & 0.351 & 0.5 & \textbf{0.352} & 0.3 & 0.346 & 0.2 \\
        T-COVID  & 0.719 & 16.0 & \textbf{0.730} & 10.1 & 0.695 & 7.5 \\
        Touche-2020  & 0.307 & 15.0 & 0.306 & 9.3 & \textbf{0.313} & 4.5 \\
        ArguAna  & 0.440 & 20.8 & 0.463 & 7.8 & \textbf{0.500}  & 4.1 \\
        C-FEVER  & \textbf{0.234} & 1375 & 0.219 & 681 & 0.213 & 332 \\
        FEVER   & \textbf{0.781} & 1584 & 0.778 & 559 & 0.764 & 264 \\
        Quora  & \textbf{0.806} & 17.5 & 0.776 & 9.2 & 0.792 & 4.5 \\
        SCIDOCS & 0.151 & 6.9 & \textbf{0.155} & 3.0 & 0.151 & 2.0 \\
        SciFact  & 0.676 & 5.7 & \textbf{0.681} & 2.4 & 0.672 & 1.4 \\
\hline 
        \textbf{Average } & \textbf{0.498} & - & 0.497 & 2.0x & 0.494 & 3.6x \\
 \hline
    \end{tabular}
    }
\vspace{-0.8em}
\end{table}

Table~\ref{tab:beir}
lists the zero-shot performance of HT  when $k=1000$ by applying the SPLADE/HT model learned from 
MS MARCO to the BEIR datasets without any additional training. 
HT$_1$ has a similar nDCG@10 score as SPLADE without HT, while having a 2x MRT speedup on average. 
HT$_3$ is even faster with 3.6x speedup, and its nDCG@10 drops in some degree to 0.494. 


\begin{figure}[htbp]
    \vspace{-1em}
\begin{center}
  \subfloat[Documents]{\rule{3cm}{0pt}}\quad\quad\quad
\subfloat[Queries]{\rule{3cm}{0pt}}\\[0.6ex]
\vspace{-0.15cm}
  \includegraphics[width=0.45\columnwidth]{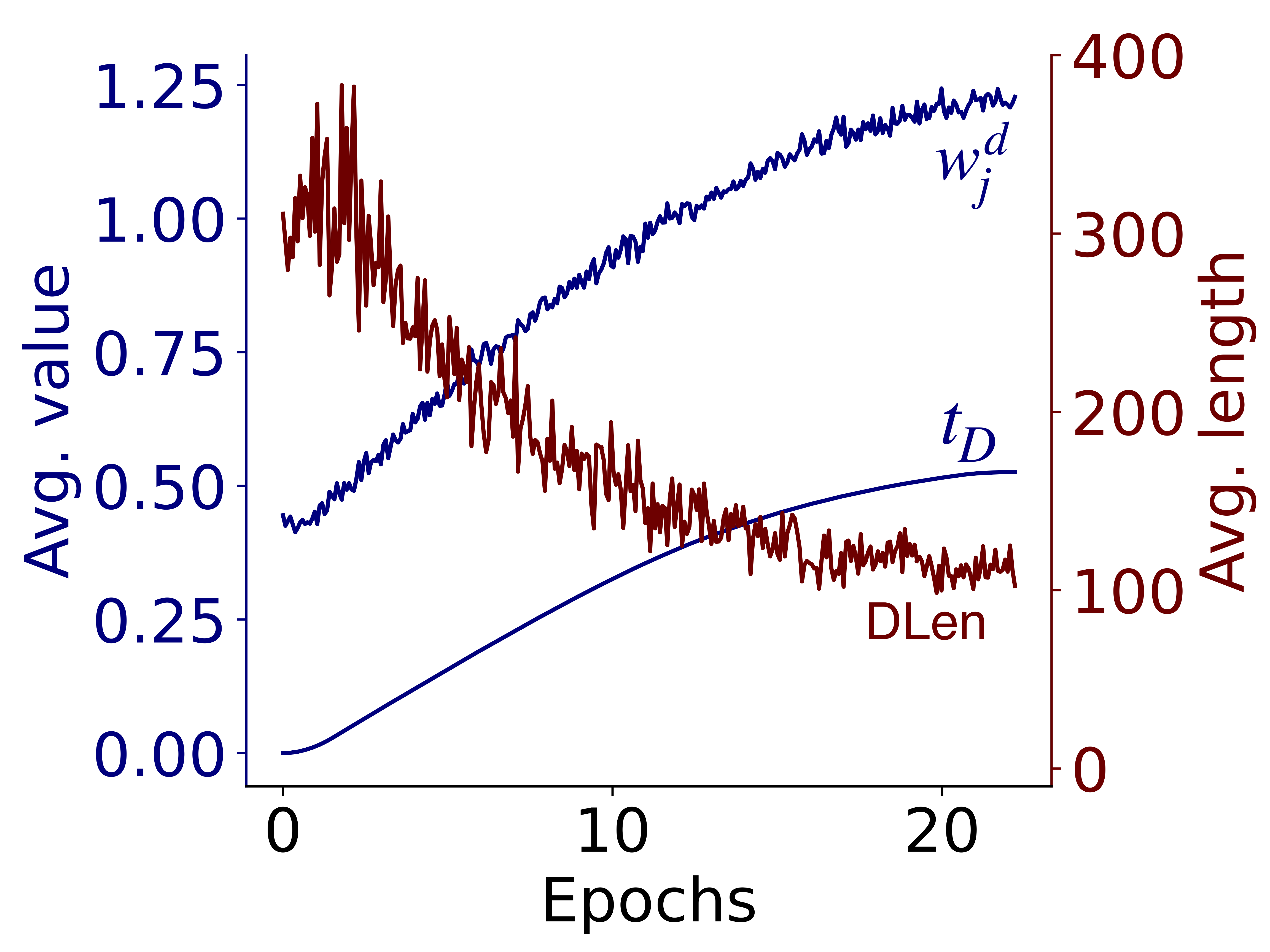}
  \includegraphics[width=0.45\columnwidth]{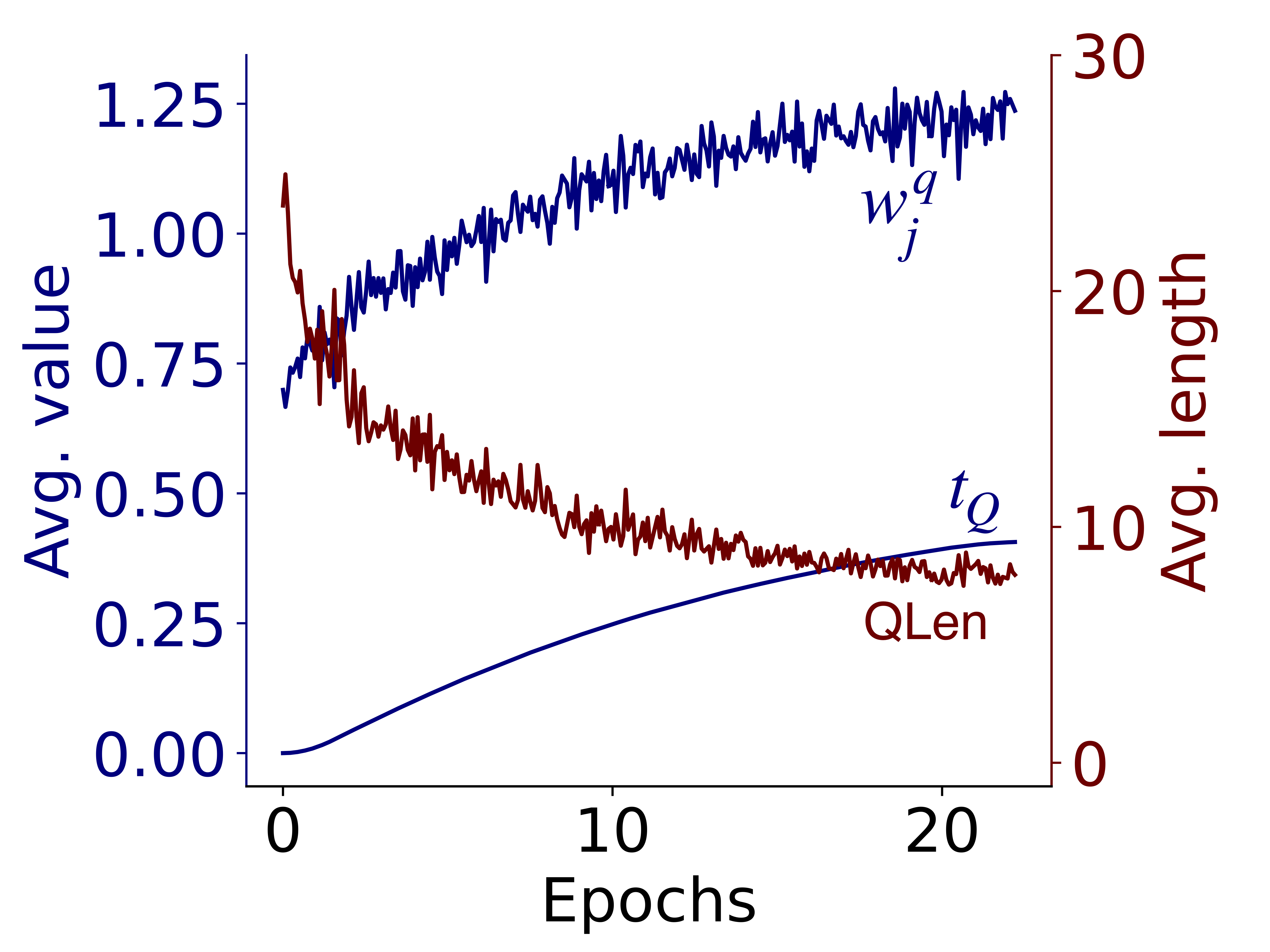}
\end{center}
    \vspace{-1em}
  \caption{  Weight/threshold/sparsity changes during training}
  \label{fig:wtandl}
\end{figure}

    \vspace{-0.2cm}

Figure \ref{fig:wtandl} depicts the average values of 
$w_j^d$, $t_D$, and $Dlen$  on the left and
$w_j^q$, $t_Q$, and $Qlen$ on the right during MS MARCO training under $HT_1$. $x$-axis is the training epoch number. 
It shows that $Dlen$ and $Qlen$ decrease while $t_D$ and $t_Q$ increase as training makes progress and 
SPLADE/HT$_1$ converges after about 20 epochs.



\textbf{Design options.} 
Table \ref{tab:softvshard} lists performance under 4 thresholding combinations 
from Row 3  to Row 7. 
$S[x]$ means 
soft thresholding function $S()$ is applied to $x$ for both training and indexing where $x$ can be documents (D) or queries (Q).
$\widehat{H}[$x$]$ means  sigmoid thresholding $\widehat{H}$ is applied in both training and indexing. 
$\widehat{H}H[x]$ means  $\widehat{H}$ is applied in training and $H$ is applied in indexing. 
$\phi[x]$ means  no thresholding is applied to $x$ during training and indexing.
When thresholding is not applied to queries,
$\widehat{H}H[D]$ 
is 1.3x faster than $S[D]$
when  $k=10$ and $k=1000$ while their  relevance scores are  similar. 
Shifting of document weight distribution by soft thresholding significantly affects retrieval time.
Rows 6 and  7 fix $\widehat{H}H[D]$ setting, and  
show that  soft thresholding is more effective in relevance than hard thresholding for query tokens. 
Shifting of query weight distribution has less effect on latency while gaining more relevance through model consistency between
training and indexing.

\begin{table}[tpbh]
\small
    \centering
        \caption{Impact of design options. MS MARCO passages.}
\label{tab:softvshard}
   \vspace{-0.2cm}
\resizebox{0.95\columnwidth}{!}{%
    \begin{tabular}{l|ccc|rr}
    \hline
     \textbf{HT Config.} & \textbf{MRR} & \textbf{MRT($P_{99}$)} & \textbf{MRT($P_{99}$)} & \textbf{Qlen} & \textbf{Dlen} \\ 
 $\lambda_T=1$     &  & $k=10$ & $k=1000$ &  &  \\ 
     \hline 
     \multicolumn{6}{l}{Soft vs. hard thresholding in 4 combinations. Fix $K=25$.  } \\ \hline
    $\phi[Q],S[D]$ & 0.3941 & 31.7(157) & 91.5(315) & 14.3 & 145 \\
    $\phi[Q],\widehat{H}H[D]$ & 0.3942 & 24.1(111) & 70.7(219) & 13.5 & 142 \\
    $S[Q], \widehat{H}H[D]$ & \textbf{0.3955} & \textbf{22.8(108)} & \textbf{62.3}(195) & 11.3 & 140\\
    $\widehat{H}H[Q], \widehat{H}H[D]$ & 0.3904 & 24.9(106) & 62.6(\textbf{182}) & 9.0 & 142 \\ \hline
    \multicolumn{6}{l}{Vary $K$. $\widehat{H}
[D]$ vs. $\widehat{H}H[D]$. Fix $S[Q]$.} 
\\ \hline
    $\widehat{H}H[D]$, $K=2.5$ & 0.3947 & 22.8(110) & 62.6(199)  & 11.5 & 149 \\
    $\widehat{H}[D]$, $K=2.5$ & \textbf{0.3963} & 41.4(198) & 112(358) & 11.5 & 421  \\
    $\widehat{H}H[D]$, $K=25$ & 0.3955 & 22.8(108) & 62.3(195) & 11.3 & 140 \\
    $\widehat{H}[D]$, $K=25$ & 0.3961 & 28.7(136) & 76.9(239) & 11.3 & 208 \\
    $\widehat{H}H[D]$, $K=250$ & 0.3946 & \textbf{21.9(102)} &\textbf{ 60.5(189)} & 11.2 & 135 \\
    $\widehat{H}[D]$, $K=250$ & 0.3947 & 23.1(112) & 63.9(203) & 11.2 & 159 \\ \hline

    \multicolumn{6}{l}{Usefulness of $L_Q$ and $L_D$. Fix $S[Q], \widehat{H}H[D]$, and $K=25$. } \\ \hline
    w/o $L_Q$ & 0.3956 & 56.2(245) & 166(502) & 20.1 & 138 \\
    w/o $L_Q$, $L_D$ & 0.3954 & 99.4(434) & 254(772) & 25.9 & 421 \\
     \hline
    \end{tabular}
    }

    \vspace{-0.3cm}
\end{table}


\textbf{Hyperparameter $K$ in sigmoid thresholding
$\widehat{H}$.} 
Table \ref{tab:softvshard} compares 
$\widehat{H}H[D]$ with 
$\widehat{H}[D]$ 
when varying $K$ from Row 8  to Row 14. 
In these cases, training always uses $\widehat{H}$ while indexing uses $\widehat{H}$  or $H$.
When $K$ is small as 2.5, 
applying $\widehat{H}$ to both training and indexing yields good relevance, but 
retrieval is about 1.8x slower 
because much more non-zero weights are kept in the index.
When $K$ becomes large as 250, training does not converge to the global optimum due to large update sizes, 
resulting in an MRR score lower than $K$=25 even with no index approximation. 
$K=25$  has a reasonable MRR while $\widehat{H}H[D]$ is up-to 26\% faster than 
$\widehat{H}[D]$.

\textbf{Retaining $L_Q$ and $L_D$.} 
Last three rows  of Table \ref{tab:softvshard} shows  that the query length is higher when $L_Q$ is removed
from the loss function, and documents get longer when   $L_D$ is removed further.
The result means $L_Q$ and $L_D$ are useful in sparsity control together with $L_T$.








            
\section{Concluding Remarks}

Our evaluation shows that  
learnable hybrid thresholding with index approximation  can effectively increase 
the sparsity of inverted indices  with 2-3x  faster retrieval and  competitive or slightly degraded relevance (0.28\% - 0.6\% MRR@10 drop).
Its trainability  allows relevance and sparsity guided threshold learning and 
it can outperform index pruning without such an optimization.
Our scheme retains a non-uniform number of non-zero token weights per vector based on a trainable
weight and threshold difference
for flexibility in relevance optimization.
Our analysis shows that  hyperparameter $K$ in sigmoid thresholding
needs to be chosen
judiciously for a small index approximation error without hurting training stability.

If a small relevance tradeoff is allowed, more retrieval time reduction is possible
when applying other related orthogonal efficiency optimization techniques
~\cite{lassance2022efficiency, mallia2022faster,Qiao20222GT,20232GT,2015ICTIR-anytime-ranking-Lin,2022ACMTrans-AnytimeRank-Mackenzie}.
\comments{
For example,  applying BM25-guided index traversal~\cite{mallia2022faster, 20232GT}
with our scheme, the response time can decrease further by around 50\%. For $k$=1000, the response time of HT$_1$ is reduced from 62.3ms to 27.6ms, and the MRR@10 is 0.3959; the response time of HT$_3$ is reduced from 40.6ms to 19.3ms, with 0.3942 MRR@10.
}
Applying hybrid thresholding HT$_{3}$ to a checkpoint of a recent efficiency-driven SPLADE 
model~\cite{lassance2022efficiency} with 0.3799 MRR@10 
on the MS MARCO passage Dev set, decreases 
the response time from 36.6ms to 21.7ms (1.7x faster)
when $k$=1000 while having 0.3868 MRR@10.
This latency can be further reduced to 14.2ms with the same MRR@10 number (0.3868) when 2GTI~\cite{20232GT} is applied to the above index.

A future study is to investigate the use of the proposed hybrid thresholding scheme
for  other learned sparse models~\cite{Mallia2021deepimpact,Lin2021unicoil,2021NAACL-Gao-COIL}. 

{\bf Acknowledgments}. We thank  Wentai Xie and anonymous referees for their valuable comments and/or help. 
This work is supported in part by NSF IIS-2225942
and has used computing resource of  NSF's ACCESS program.
Any opinions, findings, conclusions or recommendations expressed in this material
are those of the authors and do not necessarily reflect the views of the NSF.

\balance
\bibliographystyle{ACM-Reference-Format}
\normalsize
\bibliography{bib/thres.bib,bib/2022extra.bib,bib/2022refer.bib,bib/jinjin_thesis.bib,bib/mise.bib,bib/ranking.bib,bib/reference.bib,bib/url.bib}
\end{document}